\documentclass{aastex}%
\usepackage{amsmath}
\usepackage{amssymb}
\usepackage{amsfonts}
\usepackage{subfigure}
\usepackage{graphicx}%
\begin{document}
{\Large Ponderomotive modification of multicomponent magnetospheric plasma due
to electromagnetic ion cyclotron waves}

{\Large \bigskip}

\ \ \ \ \ \ \ \ \ \ \ \ \ \ \ \ \ \ \ \ \ \ \ \ \ \ \ \ \ \ \ \ \ \ \ \ \ \ \ A.
K. Nekrasov* and F. Z. Feygin**

\begin{center}
Institute of Physics of the Earth, Russian Academy of Sciences, 123995 Moscow, Russia

*) anekrasov@ifz.ru, **) feygin@ifz.ru

\bigskip
\end{center}

\textbf{Abstract }We derive the expression for the ponderomotive force in the
real multicomponent magnetospheric plasma containing heavy ions. The
ponderomotive force considered includes the induced magnetic moment of all the
species and arises due to inhomogeneity of the traveling low-frequency
electromagnetic wave amplitude in the nonuniform medium. The nonlinear
stationary force balance equation is obtained taking into account the
gravitational and centrifugal forces for the plasma consisting of the
electrons, protons and heavy ions (He$^{+}$). The background geomagnetic field
is taken for the dayside of the magnetosphere, where the magnetic field have
magnetic "holes" (Antonova and Shabansky 1968). The balance equation is solved
numerically to obtain the nonlinear density distribution of ions (H$^{+}$) in
the presence of heavy ions (He$^{+}$). It is shown that for frequencies less
than the helium gyrofrequency at the equator the nonlinear plasma density
perturbations are peaked in the vicinity of the equator due to the action of
the ponderomotive force. A comparison of the cases of the dipole and dayside
magnetosphere is provided. It is obtained that the presence of heavy ions
leads to decrease of the proton density modification.

\textbf{Keywords} Earth \textperiodcentered\ Magnetic fields
\textperiodcentered\ Plasmas \textperiodcentered\ Ponderomotive force
\textperiodcentered\ Waves

\begin{center}
\bigskip
\end{center}

\ \ \ \ \ \ 

\section{\textbf{Introduction}}

A considerable attention has been paid to study an influence of heavy ions
(mainly helium and oxygen) on the generation and dynamics of electromagnetic
ion cyclotron (EMIC) waves in the frequency range 0.5 to 5.0 Hz traveling
along the geomagnetic field lines. Using the searchcoil magnetometer and
particle spectrometer data on board of the GEOS 1 and ATS 6 satellites, Young
et al. (1981), Mauk et al. (1981) and Fraser et al. (1982) came to the
conclusion that EMIC waves are strongly controlled by the dynamics of heavy
ions. Later on, Kozyra et al. (1984) and Fraser et al. (1992) obtained similar
results, using the ISEE 1 and 2 data.

The propagation of the EMIC\ waves along magnetic field lines in a heavy ion
rich plasma is characterized by the reverse of polarization and the splitting
of the wave spectrum into two branches, the high-frequency branch
$\omega>\omega_{ch}$ ($\omega$ is the wave frequency, $\omega_{ch}$ is the
heavy ion gyrofrequency) and the low-frequency branch $\omega<\omega_{ch}$.
The two branches are separated by a stop-band. Observations on board of the
GEOS 1 and 2 satellites (Young et al. 1981) have shown that the EMIC wave
spectra are concentrated in the vicinity of the equatorial He$^{+}%
$\ gyrofrequency. These observations also showed that there is an inverse
connection between the increase of the He$^{+}$ concentration and the
appearance of the Pc1 events on the ground that confirms the important role of
the He$^{+}$ ions for the generation and propagation of the EMIC waves.

As is well-known, ponderomotive forces induced by the EMIC waves contribute to
the plasma balance in the Earth's magnetosphere (e.g. Allan 1992; Guglielmi et
al. 1993, 1995; Guglielmi and Pokhotelov 1994;  Witt et al. 1995; Pokhotelov
et al. 1996; Allan and Manuel 1996; Feygin et al. 1998; Nekrasov and Feygin
2011, 2012). In studies mentioned above, the magnetosphere has been assumed to
contain only one ion species (H$^{+}$). In this paper, we explore the effect
of the ponderomotive force in the real multicomponent magnetospheric plasma.
In the numerical analysis, the plasma consisting of the electrons, protons and
heavy ions (He$^{+}$) is considered. The background geomagnetic field is taken
for the dayside of the magnetosphere, where the magnetic field have magnetic
"holes" (Antonova and Shabansky 1968). The stationary nonlinear balance
equation is solved numerically to obtain the nonlinear density distribution of
ions (H$^{+}$) in the presence of heavy ions (He$^{+}$). Without heavy ions,
this problem has been treated by Nekrasov and Feygin (2012).

The paper is organized as follows. In Sect. 2, we describe the geomagnetic
field model of the dayside magnetosphere. The ponderomotive force in the
multicomponent plasma is derived in Sect. 3. The stationary force balance
equation is considered in Sect. 4. In Sect. 5, the results of numerical
calculations of the balance equation for the nonlinear plasma modification in
the curvature geomagnetic field are represented. Conclusive remarks and
discussion are given in Sect. 6.

\bigskip

\section{\textbf{Geomagnetic field model of the dayside magnetosphere}}

We here apply a model of the Earth's magnetic field given by Antonova and
Shabansky (1968). This model has been shown to be in a good agreement with
magnetic field observations by the HEOS 1 and 2 satellites in the dayside
magnetosphere (Antonova et al. 1983). According to this model, the geomagnetic
field is described by a superposition of two dipoles: the internal dipole with
the magnetic moment $M$ and the additional external dipole with the magnetic
moment $kM$ ($k$ is a constant parameter) disposed at the distance $a$
(measured in the units of the Earth's radius $R_{E}$) on the dayside of the
magnetosphere along the Earth-Sun line from the position of the original
dipole. The latter dipole models the distortion of the geomagnetic field
caused by the solar wind pressure. The model by Antonova and Shabansky (1968)
is sufficiently simple and convenient for an analysis of geophysical phenomena
in the dayside magnetosphere. We have described this model in our paper
Nekrasov and Feygin (2012). However, for convenience of reading, we also give
the main moments here.

In the spherical coordinate system, magnetic field components for the
two-dipole model by Antonova and Shabansky (1968) in the meridional
noon-midnight plane have the form%
\begin{equation}
B_{r}=-\frac{2B_{E}x}{r^{3}}\alpha,
\end{equation}%
\begin{equation}
B_{\varphi}=\frac{B_{E}}{r^{3}}\sqrt{1-x^{2}}\beta,
\end{equation}
where $x=\sin\varphi$, $\varphi$ is the geomagnetic latitude, $r$ is measured
in the units $R_{E}$, and $B_{E}$ $=0.311$ G is the equatorial magnetic field
at the Earth's surface. Coefficients $\alpha$ and $\beta$ are the following:
\begin{equation}
\ \alpha=1-\frac{kr^{3}\left(  a^{2}-2r^{2}+ar\sqrt{1-x^{2}}\right)
}{2\left(  a^{2}+r^{2}-2ar\sqrt{1-x^{2}}\right)  ^{5/2}},
\end{equation}%
\begin{equation}
\ \ \beta=1+\frac{kr^{3}\left[  \sqrt{1-x^{2}}\left(  a^{2}+r^{2}\right)
-ar\left(  2+x^{2}\right)  \right]  }{\sqrt{1-x^{2}}\left(  a^{2}%
+r^{2}-2ar\sqrt{1-x^{2}}\right)  ^{5/2}}.
\end{equation}
The total magnetic field $B$ in an arbitrary point of the field line\ can be
defined from (1) and (2) as%
\begin{equation}
B=\frac{B_{E}}{r^{3}}\left[  4x^{2}\alpha^{2}+\left(  1-x^{2}\right)
\beta^{2}\right]  ^{1/2}.
\end{equation}
An equation for the field line is determined by $dr/rd\varphi=B_{r}%
/B_{\varphi}$ or
\begin{equation}
\frac{dr}{dx}=-\frac{2xr}{1-x^{2}}\frac{\alpha}{\beta}.
\end{equation}

The magnetic field $B$ described by (5) has along the near boundary field
lines two minima located symmetrically relative to the equator (Antonova and
Shabansky 1968). When $a$ tends to infinity, the values $\alpha$ and $\beta$
tend to $1$ (see 3 and 4). In this case, we have a transition to the
one-dipole approximation. The dependence of the geomagnetic field (5) on $x$
for different distances from the Earth is shown in Fig. 1, where $L$ is the
McIlwain parameter.

\bigskip\ \ \ \ \ \ \ \ \ \ \ \ \ \ \ \ \ \ \ \ \ \ \ \ \ \ \ \ \ \ \ \ \ \ \ \ \ \ 

\section{\textbf{Ponderomotive force in the multicomponent plasma}}

\subsection{\textbf{Ponderomotive force due to the longitudinal inhomogeneity
of the wave amplitude}}

To find the ponderomotive force of electromagnetic waves in the multicomponent
plasma, we use the results of the paper by Nekrasov and Feygin (2006). The
equation for the nonlinear slow velocity $\left\langle \mathbf{v}%
_{j2}\right\rangle $ is given by
\begin{equation}
\frac{\partial\left\langle \mathbf{v}_{j2}\right\rangle }{\partial t}%
=\frac{q_{j}}{m_{j}}\left(  \left\langle \mathbf{E}_{2}\right\rangle
+\left\langle \mathbf{F}_{j2}\right\rangle \right)  ,
\end{equation}
where $q_{j}$ and $m_{j}$ are the charge and mass of the species $j$,
$\left\langle \mathbf{E}_{2}\right\rangle $ is the slow nonlinear electric
field, $\left\langle \mathbf{F}_{j2}\right\rangle $ is equal to%
\begin{equation}
\left\langle \mathbf{F}_{j2}\right\rangle =-\frac{m_{j}}{q_{j}}\left\langle
\mathbf{v}_{j1}\cdot\mathbf{\nabla v}_{j1}\right\rangle +\frac{1}%
{c}\left\langle \mathbf{v}_{j1}\times\mathbf{B}_{1}\right\rangle ,
\end{equation}
the angle brackets $\left\langle {}\right\rangle $ denote the time-averaging
over fast oscillations. The subscript $1$ in (8) relates to the linear
perturbations of the velocity, $\mathbf{v}_{j1}$, and magnetic field,
$\mathbf{B}_{1}$.

We are interested in the nonlinear motion of a plasma along the background
magnetic field. We assume that the latter has the $\mathbf{z}$-direction. In
the case of the EMIC waves also traveling along the magnetic field, the
equation (7) for the longitudinal velocity $\left\langle v_{j2z}\right\rangle
$ takes the form%
\begin{equation}
\frac{\partial\left\langle v_{j2z}\right\rangle }{\partial t}=-\frac{q_{j}%
}{m_{j}}\frac{1}{\omega_{pe}^{2}}\sum_{l}\omega_{pl}^{2}\left(  \left\langle
F_{l2z}\right\rangle -\left\langle F_{j2z}\right\rangle \right)
\end{equation}
and the value $\left\langle F_{j2z}\right\rangle $ defined by (8) is given by
\begin{equation}
\left\langle F_{j2z}\right\rangle =-\frac{q_{j}}{2m_{j}}\frac{1}{\sigma
\omega\left(  \sigma\omega-\omega_{cj}\right)  }\frac{\partial\left\langle
\mathbf{E}_{1}^{2}\right\rangle }{\partial z},
\end{equation}
where $\omega>0$ is the wave frequency, $\sigma=\pm1$ denotes the left ($+$)
or right ($-$) polarization of the wave, $\omega_{pj}=\left(  4\pi n_{j}%
q_{j}^{2}/m_{j}\right)  ^{1/2}$ is the plasma frequency, $n_{j}$ is the
background number density, $\omega_{cj}=q_{j}B/m_{j}c$ is the cyclotron
frequency, $c$ is the speed of light in vacuum and $\mathbf{E}_{1}%
=\operatorname{Re}\mathbf{E}_{10}\exp\left(  i\int^{z}kdz-i\omega t\right)  $
is the wave electric field with the nonuniform amplitude. When obtaining (9)
and (10), we have used (11)-(13), (16) and (17) in Nekrasov and Feygin (2006),
have taken into account that $v_{j1z}=0$ for waves under consideration and
have assumed that $\partial/\partial z\gg k\partial/\omega\partial t$, where
$\partial/\partial t$ is applied to the wave amplitude. The right-hand side of
(9) (without $m_{j}$) is the particular (see below) ponderomotive force acting
on species $j$.

Using (9), we can define the total particular ponderomotive force in the
multicomponent plasma as
\begin{equation}
F_{p1z}=\sum_{j}m_{j}n_{j}\frac{\partial\left\langle v_{j2z}\right\rangle
}{\partial t}=-\frac{1}{\omega_{pe}^{2}}\sum_{j,l}q_{j}n_{j}\omega_{pl}%
^{2}\left(  \left\langle F_{l2z}\right\rangle -\left\langle F_{j2z}%
\right\rangle \right)  .
\end{equation}
Taking into account (10), the expression (11) gets the following:%
\begin{equation}
F_{p1z}=-\frac{1}{8\pi}\sum_{j\neq e}\frac{\omega_{pj}^{2}}{\omega_{cj}\left(
\sigma\omega-\omega_{cj}\right)  }\frac{\partial\left\langle \mathbf{E}%
_{1}^{2}\right\rangle }{\partial z},
\end{equation}
where we have neglected small terms proportional to $m_{e}$, which describe
the ponderomotive force acting on the electrons.

\bigskip

\subsection{\textbf{Ponderomotive force due to the transverse inhomogeneity of
the wave amplitude. Magnetic moment}}

It is known that if the medium has the own magnetic moment (magnetic field)
$\mathbf{M}$ and is embedded in the external nonuniform magnetic field
$\mathbf{B}$, then this medium is subjected by the action of the force
\begin{equation}
\mathbf{F}_{\mathbf{M}}=\left(  \mathbf{M\cdot\nabla}\right)  \mathbf{B}%
\end{equation}
(see e.g. Landau and Lifshits 1982). The EMIC waves propagating along the
geomagnetic field generate the nonlinear magnetic field that can be found from
(3), (16) and (17) in (Nekrasov and Feygin 2006). Neglecting in (16) (Nekrasov
and Feygin 2006) small terms and passing to the space-time representation, we
find the equation for the transverse nonlinear electric field $\left\langle
E_{2x}\right\rangle $
\begin{equation}
\left(  \mathbf{\nabla}^{2}c_{A}^{2}-\frac{\partial^{2}}{\partial t^{2}%
}\right)  \left\langle E_{2x}\right\rangle =\frac{c_{A}^{2}}{c^{2}}%
\frac{\partial^{2}}{\partial t^{2}}\left\langle C_{x}\right\rangle ,
\end{equation}
where $\mathbf{\nabla}^{2}=\partial^{2}/\partial y^{2}+\partial^{2}/\partial
z^{2}$, $c^{2}/c_{A}^{2}=\sum_{j}\omega_{pj}^{2}/\omega_{cj}^{2}$, $c_{A}$ is
the Alfv\'{e}n velocity and
\begin{equation}
\frac{\partial\left\langle C_{x}\right\rangle }{\partial t}=-\frac{c}{2B}%
\sum_{j}\frac{\omega_{pj}^{2}}{\left(  \sigma\omega-\omega_{cj}\right)  ^{2}%
}\frac{\partial\left\langle \mathbf{E}_{1}^{2}\right\rangle }{\partial y}%
\end{equation}
(see (17) in (Nekrasov and Feygin 2006)).

We further assume that $\partial^{2}/\partial y^{2}\gg\partial^{2}/\partial
z^{2}$ and $c_{A}^{2}\partial^{2}/\partial y^{2}\gg\partial^{2}/\partial
t^{2}$. Then from (14) and (15), we obtain%
\begin{equation}
\frac{\partial\left\langle E_{2x}\right\rangle }{\partial y}=-\frac{1}%
{2cB}\sum_{j}\frac{\omega_{pj}^{2}}{\left(  \sigma\omega-\omega_{cj}\right)
^{2}}\frac{\partial\left\langle \mathbf{E}_{1}^{2}\right\rangle }{\partial t}.
\end{equation}
From Faraday's equation%
\begin{equation}
\mathbf{\nabla\times}\left\langle \mathbf{E}_{2}\right\rangle =-\frac{1}%
{c}\frac{\partial\left\langle \mathbf{B}_{2}\right\rangle }{\partial t},
\end{equation}
we can find the induced nonlinear magnetic field $\left\langle \mathbf{B}%
_{2}\right\rangle $. Substituting (17) into (16), we have in our case for the
multicomponent plasma%
\begin{equation}
\left\langle B_{2z}\right\rangle =-\frac{1}{2B}\sum_{j}\frac{\omega_{pj}^{2}%
}{\left(  \sigma\omega-\omega_{cj}\right)  ^{2}}\left\langle \mathbf{E}%
_{1}^{2}\right\rangle .
\end{equation}
We note that this expression for the two-component plasma (electrons and ions)
has been derived by Nekrasov and Feygin (2005).

The magnetic moment $\mathbf{M}$ is connected with the induced magnetic field
as $4\pi\mathbf{M=}\left\langle \mathbf{B}_{2}\right\rangle $. Using this
relation, we obtain in our case the expression (13) for $\mathbf{F}%
_{\mathbf{M}}$ along the magnetic field
\begin{equation}
F_{\mathbf{M}z}=\frac{1}{4\pi}\left\langle B_{2z}\right\rangle \frac{\partial
B}{\partial z}.
\end{equation}
This expression for the two-component plasma has also been obtained by
Nekrasov and Feygin (2005), but in a different way. However, we see here under
which conditions (18) takes place.

\bigskip

\subsection{\textbf{Total ponderomotive force in the nonuniform medium}}

The total ponderomotive force $F_{pz}$ is equal to sum of (12) and (19)%
\begin{equation}
F_{pz}=-\frac{1}{8\pi}\sum_{j\neq e}\frac{\omega_{pj}^{2}}{\omega_{cj}\left(
\sigma\omega-\omega_{cj}\right)  }\left[  \frac{\partial\left\langle
\mathbf{E}_{1}^{2}\right\rangle }{\partial z}+\frac{\omega_{cj}}{\left(
\sigma\omega-\omega_{cj}\right)  }\frac{\partial B}{B\partial z}\left\langle
\mathbf{E}_{1}^{2}\right\rangle \right]  ,
\end{equation}
where we have neglected the contribution of the electrons into (18). We
further assume that the wave amplitude depends on the $z$-coordinate because
of the longitudinal medium inhomogeneity. In this case, the amplitude of the
wave $E_{10}$ in the WKB-approximation is proportional to $N_{z}^{-1/2}$,
where $N_{z}$ is the refractive index equal to
\begin{equation}
N_{z}^{2}=1-\sum_{j\neq e}\frac{\omega_{pj}^{2}}{\omega_{cj}\left(
\sigma\omega-\omega_{cj}\right)  }%
\end{equation}
(see e.g. (14) in (Nekrasov and Feygin 2006)). When obtaining (21), we have
used the condition of quasineutrality $\sum_{j}q_{j}n_{j}=0$. We see that the
first term in the square brackets in (20) can be written in the form%
\[
F_{p1z}=\frac{1}{8\pi}\left(  N_{z}^{2}-1\right)  \frac{\partial\left\langle
\mathbf{E}_{1}^{2}\right\rangle }{\partial z}.
\]
Calculating the value $\partial\left\langle \mathbf{E}_{1}^{2}\right\rangle
/\partial z$ through $\partial N_{z}^{2}/\partial z$, using (21) and
substituting the result into (20), we find%
\begin{equation}
F_{pz}=-\frac{\left\langle \mathbf{E}_{1}^{2}\right\rangle }{16\pi}\sum_{j\neq
e}\frac{\omega_{pj}^{2}}{\omega_{cj}\left(  \omega_{cj}-\sigma\omega\right)
}\left\{  \frac{\partial\ln n_{j}}{\partial z}+\left[  \frac{\sigma\omega
}{\omega_{cj}-\sigma\omega}+\frac{2\omega_{cj}}{\left(  \omega_{cj}%
-\sigma\omega\right)  }\frac{1}{N_{z}^{2}}\right]  \frac{\partial\ln
B}{\partial z}\right\}  .
\end{equation}
Thus, the ponderomotive force in the multi-component plasma is the sum of the
ponderomotive forces for each ion species.

\bigskip

\section{\textbf{Stationary force balance equation }}

From equations of motion for the species $j$ in the second approximation on
the wave amplitude averaged over fast oscillations, we can obtain the force
balance equation in the stationary state. The total equation of motion
contains except the electromagnetic force also the gradient of thermal
pressure, the gravity and centrifugal force (e.g. Lemaire 1989; Persoon et al. 2009).

We assume for simplicity that all the species have the equal temperatures
$T_{j}=T=const$. Consider the three-component plasma consisting of the
electrons ($e$), protons ($i$) and heavy ions ($h$). In the stationary case,
adding the corresponding equations of motion for each species, we obtain for
the nonlinear time-averaging density perturbations $n_{j2}$ the following
force balance equation along the magnetic field line:%
\begin{equation}
T\nabla_{\Vert}\left[  2n_{i2}+\left(  1+\frac{q_{h}}{q_{i}}\right)
n_{h2}\right]  =\left(  m_{i}n_{i2}+m_{h}n_{h2}\right)  \left(  g_{\Vert
}\mathbf{+}n_{\Omega\Vert}\Omega^{2}R_{E}r\cos\varphi\right)  +F_{p\Vert},
\end{equation}
where the subscript $\Vert$ denotes the local $\mathbf{z}$-direction. Here, we
have used the condition of quasineutrality and the equality $q_{i}=-q_{e}$. In
(23), the value $g_{\Vert}=\mathbf{g\cdot b=-}g_{E}B_{r}/r^{2}B$ is the
longitudinal gravitational acceleration ($g_{E}=9.8$ m sec$^{-2}$), $\Omega$
is the Earth's rotation frequency and $n_{\Omega\parallel}=\mathbf{n}_{\Omega
}\cdot\mathbf{b}$, where $\mathbf{n}_{\Omega}$ and $\mathbf{b}$ are the unit
vectors along the centrifugal force and magnetic field, respectively.The value
$n_{\Omega\parallel}$ is equal to $n_{\Omega\parallel}=\left(  1+\beta
/2\alpha\right)  \left(  B_{r}/B\right)  \cos\varphi$ (see Nekrasov and Feygin
2012). We note that in the last paper in (14) $\cos^{2}\varphi$ stands
erroneously instead of $\cos\varphi$.

In (22), we will express the value $\left\langle \mathbf{E}_{1}^{2}%
\right\rangle $ through the amplitude of the wave magnetic field at the
equator $B_{10}$. From Faraday's equation, it is followed that $N_{z}%
^{2}\left\langle \mathbf{E}_{1}^{2}\right\rangle =\left\langle \mathbf{B}%
_{1}^{2}\right\rangle $, where $\left\langle \mathbf{E}_{1}^{2}\right\rangle
\propto N_{z}^{-1}$. Thus, we obtain that $\left\langle \mathbf{E}_{1}%
^{2}\right\rangle =B_{10}^{2}\left(  N_{z0}N_{z}\right)  ^{-1}$ for
circularly-polarized waves. Here and below, the subscript $0$ relates to the
values at the equator. The operator $\nabla_{\parallel}$ in (23) is defined by
the relation $\nabla_{\parallel}=\mathbf{b}\cdot\mathbf{\nabla}$ and has been
found in (Nekrasov and Feygin 2012)
\begin{equation}
\nabla_{\parallel}=2R_{E}^{-1}\eta^{-1/2}\frac{d}{dx},
\end{equation}
where%

\begin{equation}
\eta=\left(  \frac{dr}{dx}\right)  ^{2}+\frac{r^{2}}{1-x^{2}}.
\end{equation}

We further connect the perturbation $n_{h2}$ with $n_{i2}$. From the nonlinear
continuity equation and equation of motion (9), we can obtain an estimation of
this connection
\begin{equation}
n_{h2}\sim n_{i2}\frac{m_{i}}{m_{h}}\frac{H}{P}\frac{\left(  1-\nu_{i0}%
\frac{B_{0}}{B}\right)  }{\left(  1-\nu_{h0}\frac{B_{0}}{B}\right)  },
\end{equation}
where $P=\rho_{i}/\rho_{i0},H=\rho_{h}/\rho_{i0},\rho_{i}=n_{i}m_{i},\rho
_{h}=m_{h}n_{h},\nu_{i0}=\sigma\omega/\omega_{ci0},\nu_{h0}=\sigma
\omega/\omega_{ch0}$.

For the three-component plasma, substituting (22), (24) and (26) into (23), we
obtain the following equation:%
\begin{equation}
\frac{d\delta}{dx}=\frac{1}{\lambda}\left(  A_{1}-\frac{d\lambda}{dx}\right)
\delta+\frac{1}{\lambda}A_{2}\left(  A_{3}+A_{4}+A_{5}+A_{6}\right)  ,
\end{equation}
where $\delta=\rho_{i2}/\rho_{i0}$, $\rho_{i2}=m_{i}n_{i2}$ and%
\begin{equation}
\lambda=2+\left(  1+\frac{q_{h}}{q_{i}}\right)  \frac{m_{i}}{m_{h}}\frac{H}%
{P}\frac{\left(  1-\nu_{i0}\frac{B_{0}}{B}\right)  }{\left(  1-\nu_{h0}%
\frac{B_{0}}{B}\right)  }.
\end{equation}
In (27), we have introduced the notations%
\begin{align}
A_{1}  &  =\frac{2R_{E}x\alpha\mu g_{eff}}{\left(  1-x^{2}\right)  \left\vert
\beta\right\vert c_{si}^{2}r},\\
A_{2}  &  =-\frac{B_{10}^{2}}{8\pi\rho_{i0}c_{si}^{2}}\frac{B_{0}}{B}\left(
\frac{1}{\mu\mu_{0}}\right)  ^{1/2}\frac{\left(  1-\nu_{i0}\right)  ^{1/2}%
}{\left(  1-\nu_{i0}\frac{B_{0}}{B}\right)  ^{1/2}},\nonumber\\
A_{3}  &  =\frac{1}{P^{1/2}}\frac{dP}{dx},\nonumber\\
A_{4}  &  =\frac{\left(  1-\nu_{i0}\frac{B_{0}}{B}\right)  }{\left(
1-\nu_{h0}\frac{B_{0}}{B}\right)  }\frac{1}{P^{1/2}}\frac{dH}{dx},\nonumber\\
A_{5}  &  =P^{1/2}\frac{\left(  \nu_{i0}\frac{B_{0}}{B}+\frac{2}{N_{z}^{2}%
}\right)  }{\left(  1-\nu_{i0}\frac{B_{0}}{B}\right)  }\frac{1}{B}%
\frac{\partial B}{\partial x},\nonumber\\
A_{6}  &  =\frac{H}{P^{1/2}}\frac{\left(  1-\nu_{i0}\frac{B_{0}}{B}\right)
}{\left(  1-\nu_{h0}\frac{B_{0}}{B}\right)  ^{2}}\left(  \nu_{h0}\frac{B_{0}%
}{B}+\frac{2}{N_{z}^{2}}\right)  \frac{1}{B}\frac{\partial B}{\partial
x},\nonumber
\end{align}
where $c_{si}^{2}=2T/m_{i}$ and%
\begin{equation}
g_{eff}=g_{E}+g_{cf}=g_{E}-\left(  1+\beta/2\alpha\right)  \Omega^{2}%
R_{E}r^{3}\cos^{2}\varphi,
\end{equation}%
\begin{align}
\mu &  =1+\frac{H}{P}\frac{\left(  1-\nu_{i0}\frac{B_{0}}{B}\right)  }{\left(
1-\nu_{h0}\frac{B_{0}}{B}\right)  },\\
N_{z}^{2}  &  =4\pi c^{2}\frac{\rho_{i}}{B^{2}}\frac{1}{\left(  1-\nu
_{i0}\frac{B_{0}}{B}\right)  }\mu.\nonumber
\end{align}
We consider that $N_{z}^{2}\gg1$. The value $\mu_{0}$ is obtained from (31) at
$x=0$. When obtaining $A_{1}$, we have used (1), (5), (6) and (25).

For the equilibrium mass density of H$^{+}$ ($i$) and He$^{+}$ ($h$), we take
the power law form to describe the longitudinal field line distribution,
$\rho_{j}\propto r^{-\gamma}$, where $j=i,h$. For the large distances from the
Earth's surface which we consider below, the choice $\gamma=1$ is the best one
to be appropriate to experimental data (Denton et al. 2006). Thus, we set%
\[
\rho_{j}\left(  x\right)  =\rho_{j0}\left(  1-x^{2}\right)  ^{-1}.
\]
This formula can be applied up to $\varphi\approx\pm50-60^{0}$ (Denton et al.
2006). In this case, $P=\left(  1-x^{2}\right)  ^{-1}$ and $H=H_{0}\left(
1-x^{2}\right)  ^{-1}$. These values will be substituted to (28) and (29). We
note that in our case $m_{h}=4m_{i}$.

\bigskip

\section{\textbf{Numerical analysis}}

Equation (27) for the two-component magnetospheric plasma has been analyzed in
(Nekrasov and Feygin 2012) by the Runge-Kutta method, making use of the
boundary condition\textbf{\ }$\delta=0.06$\textbf{\ }at\textbf{\ }$\ x=0$.
Thus, we have obtained effects of the action of the ponderomotive,
gravitational and centrifugal forces on the plasma density distribution at
$x\neq0$, at the same time perturbations at the equator remained unchanged. In
this paper, we set the boundary condition $\delta=0$ at $x=0.8$. This case
permits us to consider the plasma density modification in the region of the
equator. For numerical calculations, we have used the same parameters for the
magnetic field model as that in (Nekrasov and Feygin 2012): $a=33$ and $k=13$
(except for Figs. 3a,b). These parameters correspond to the dayside boundary
of the magnetosphere at the distance $10R_{E}$ obtained by HEOS 1 and 2
satellites (Antonova at al. 1983). Other parameters are the following:
$c_{si}^{2}=10^{9}$ m$^{2}$ $\sec^{-2}$, $\rho_{i0}=1.67\times10^{-20}$ kg
m$^{-3}$ for all $L$ near the midday boundary of the Earth's magnetosphere,
where the plasma density depends weakly on $L$ (Chappel 1974; Carpenter and
Anderson 1992). In all the numerical calculations, we have taken $\sigma=+1$
and assumed that $\nu_{h0}<1$ to avoid the singularity $1-\nu_{h}=0$. As heavy
ions, we use He$^{+}$. \textbf{ }

Figures 2 depict the distribution of the normalized nonlinear proton density
$\delta=\rho_{i2}/\rho_{i0}$ along the field line for different values of
$H_{0}=0,0.5,1.0$ at $L=6,10$ for $B_{10}=10^{-5}$ G. We have set $\nu
_{i0}=0.1$ to have $\nu_{h0}<1$. We see that the plasma density perturbation
increases in the direction of the equator due to the action of the
ponderomotive force (without the latter see Fig. 5). The equatorial value of
$\delta$ decreases with increasing of $H_{0}$ because of increasing of
$\lambda$ and $\mu$ with $H_{0}$ (see (27)-(29) and (31)).

Figures 3 represent a difference in the $\delta$-distribution for the dipole
($k=0$) and two-dipole ($k=13$) geomagnetic field at $L=8$ and $H_{0}=0,1.0$
for $B_{10}=10^{-5}$ G. We see that the peak value of $\delta$ at the equator
is smaller for $k=13$ than for $k=0$. This can be explained by increasing of
the equatorial geomagnetic field $B$ due to $k\neq0$ (see (3)-(5)), which
stands in the denominator of $A_{2}$ (see (29)) and decreases the
ponderomotive force. The presence of heavy ions also results in a decreased
influence of the ponderomotive force on the ions due to increase of the
parameters $\lambda$ and $\mu$ (as in Fig. 2). Therefore, for $H_{0}=1.0$, the
values of $\delta$ are smaller than for $H_{0}=0$. We note that the curves for
$H_{0}=0.5$ are here and in Figs. 4 situated between the curves with
$H_{0}=0,1.0$ (except Fig.5).

Figures 4 show the dependence of the relative density perturbation $\delta$ on
$x$ for different values of the wave amplitude $B_{10}=(1,3,5)\times10^{-5}$ G
for $L=10$ and $H_{0}=0,1.0$. It can be seen that increased wave-amplitude
yields increased values of $\delta$ along field lines. The smaller values of
$\delta$ at $H_{0}=1.0$ are explained in the description of Figs. 2a, 2b and
3a, 3b.

Figure 5 describes formally the case $B_{10}=0$ to show the role of the
gravitational and centrifugal forces (see (30)). It is followed from (27) that
gravitational force increases the plasma density (not only perturbations!)
with larger distances from the equator. This is shown by the curve 1. The
centrifugal force has the opposite direction (see (30)) and drives a plasma to
the equator. Far from equator, the gravitational force is larger than the
centrifugal one (see also Sect. 6). In the vicinity of the equator, on the
contrary, the centrifugal force is larger. This results in a peak of plasma
density at the equator. The curve 2 shows the corresponding distribution. The
equatorial values of $\delta$ for the curves 1 and 2 are decreasing with
increasing $H_{0}$. The boundary condition for the curves 1 and 2 is
$\delta=0.01$ at $x=0.8$ since to solve (27) for $A_{2}=0$ it is necessary to
have a nonzero boundary condition for $\delta$.

All the figures demonstrate that the density perturbation is peaked at the
equator due to ponderomotive and centrifugal forces. Such peaking is observed
in the magnetosphere of the Earth (Denton et al. 2006). \ 

\bigskip

\section{\textbf{Conclusion and Discussion\bigskip}}

We conclude by summarizing main results obtained in this paper:

1). We have derived the general expression for the ponderomotive force induced
by electromagnetic ion cyclotron waves in a multicomponent plasma containing
different species of ions.

2). The correct equation of the force balance along the magnetic field lines,
which contains the perturbed thermal plasma pressure together with the
ponderomotive force, having the same (second) order of magnitude, has been considered.

3). We have investigated the effect of the ponderomotive force on the
perturbed ion (proton) density distribution in the presence of helium ions.

4). It has been shown that for frequencies less than the helium gyrofrequency
at the equator the nonlinear plasma density perturbations are peaked in the
vicinity of the equator due to the action of the ponderomotive force. The
maximum of the ion (proton) density perturbation has been obtained to decrease
with increasing of the heavy ion (He$^{+}$) mass density.

5). We have obtained that larger wave-amplitudes inducing more ponderomotive
force result in larger $\delta$-perturbations.

Theoretical and numerical results given above in this section are the main
points of our exploration. The derivation of expression for the ponderomotive
force in the multicomponent plasma containing different species of heavy ions
is particularly important. This permits us to consider the influence of wave
perturbations on the redistrbution of magnetospheric plasma in the direction
of the equator in conditions close to real ones. As we see, the equatorial
value of the proton density distribution should drop for more abundance of
He$^{+}$.

We have considered the relative role of the gravitational and centrifugal
forces when the wave activity is\textbf{ }absent, $B_{10}=0$ (Fig. 5). We see
from (30) that the gravitational and centrifugal forces along the magnetic
field lines have opposite directions. At some latitude $\varphi_{eq}$, both
forces are equal to each other. Taking into account (6), the estimation of
$\varphi_{eq}$ for $\alpha\sim\beta\sim1$ is the following:%
\begin{equation}
\cos\varphi_{eq}\approx\left(  \frac{2g_{E}}{3\Omega^{2}R_{E}r_{0}^{3}%
}\right)  ^{1/8},
\end{equation}
where $r_{0}R_{E}$ is the equatorial distance of the given magnetic field line
from the Earth's center ($r_{0}=L$). At $\varphi>\varphi_{eq}$, the
gravitational force is larger than the centrifugal one. When $\varphi
<\varphi_{eq}$, the centrifugal force is dominant. For $r_{0}=8$, we obtain
$\cos\varphi_{eq}\approx0.89$ or $\varphi_{eq}\approx27%
{{}^\circ}%
$. When $r_{0}=10$, we have $\cos\varphi_{eq}\approx0.82$ and $\varphi
_{eq}\approx35%
{{}^\circ}%
$.

From (27), we can roughly estimate the relative contribution of the
ponderomotive force $F_{p}$ and of the gravitational and centrifugal forces
$F_{g+c}$ as
\begin{equation}
\frac{F_{p}}{F_{g+c}}\sim\frac{A_{2}A_{3}}{A_{1}\delta}.
\end{equation}
Assuming that $\delta$-perturbation is due to the wave action, we estimate in
our case $\delta\sim A_{2}$. Substitution of $\delta$ in (33) gives%
\[
\frac{F_{p}}{F_{g+c}}\sim\frac{A_{3}}{A_{1}}\approx\frac{c_{si}^{2}r_{0}%
}{R_{E}g_{E}}.
\]
For $r_{0}=10$ and $c_{si}^{2}=10^{9}$ m$^{2}$ $\sec^{-2}$, we find
$F_{p}/F_{g+c}\sim160$. We see that the ponderomotive force is dominant
because $A_{1}\ll1$ (for $x\lesssim1$). We note that a change of the boundary
condition by the nonzero valus of $\delta$ in Fig. 2b does not influence on
the form of the curves. Thus, we can expect substantial increases in $\delta
$-perturbations during more geomagnetically active periods (Figs. 4a, 4b).
This result is qualitatively consistent with observational data, such as
Denton et al. (2006), which show a larger equatorial mass density peak with
larger wave amplitudes (see their Figure 12). As a possible reason for this
effect, Denton et al. (2006) indicate a role of the ponderomotive force in
driving ions up the field line.

We have investigated the nonlinear plasma density redistribution for the
Antonova and Shabansky (1968) model of the geomagnetic field $B$ given in Fig.
1. This choice is justified by its sufficient simplicity. In addition, the
model by Antonova and Shabansky (1968) provides a complete analytical
description of the geomagnetic field on the dayside of the Earth's
magnetosphere and is in accordance with satellite measurements. The use of a
more complex geometry for $B$ such as T96 (Tsyganenko 1995) could be justified
for taking into account some details (the local time, position, the solar wind
parameters etc.). However, the results obtained will qualitatively remain the
same because the main form of the magnetic field lines does not essentially change.

In this paper, we have applied the theory of perturbations to obtain the
ponderomotive force expression and the force balance equation. We assumed that
$n_{j2}\ll n_{j0}$. The value $\delta$ for parameters used is in the region
$\delta\sim2\times10^{-3}\div6\times10^{-2}$ (see Figs. 2-4). Thus,
perturbations are small in comparison with the background density. During
active wave periods, density perturbations can be of the same order of
magnitude as background values (see Denton et al. 2006). Therefore, there is
some problem to compare theoretical \ and experimental results. However, for
the limiting estimation, we could take such wave amplitudes, for which
$\delta\sim1$, where $\delta\sim\left\vert A_{2}\right\vert $. For $\rho_{i0}$
and $c_{si}^{2}$ given in Sect. 5 and $H_{0}=1$, we obtain roughly $B_{10}%
\sim10^{-4}$ G.

We have derived the ponderomotive force for circularly-polarized waves
traveling along the magnetic field. For example, in papers by Denton et al.
(2004, 2006) and Takahashi et al. (2006), toroidal Alfv\'{e}n waves were
discussed. Therefore, it is worth to derive ponderomotive forces for other
types of perturbations. The case of the strong nonlinearity when density
perturbations are the same as background values deserves also to be examined.

We have assumed that condition of quasineutrality is satisfied in the
background and perturbation states. It is followed, for example, from the
paper by Denton et al. (2006), that equatorial peaking is observed for ions
and absent for electrons. We think that a large electric field should arise in
this case. \ 

It is obvious that for a detailed comparison of theoretical results with
observational data, theoretical models should be adequate to real experimental
conditions and observations and vice versa. It is important to be sure that a
stationary symmetric theoretical picture considered here is relevant to real
situations. The equatorial peaking in this picture is possible, if the wave
action is symmetric and simultaneous from both ionospheric boundaries to the
equator during some time for establishment of the stationary state and wave
amplitudes should decrease from the ionosphere to the equator. In other cases,
dynamical (nonstationary) processes can play a role.

\bigskip

\textbf{Acknowledgements }We gratefully thank the anonymous referee for
his/her very constructive and useful comments which have helped considerably
to improve a presentation of this paper. We also acknowledge the financial
support from the Russian Foundation for Basic Research, research grants No.
11-05-00920 and the Program of Russian Academy of Sciences No. 22 and 4.

\bigskip

\section*{\textbf{References}}

Allan, W.\textbf{:} J. Geophys. Res. \textbf{97,} 8483 (1992)\ \ 

Allan, W., Manuel, J. R.: Ann. Geophys. \textbf{14,} 893 (1996)

Antonova, A.E., Shabansky, V.P.: Geomagn. Aeron. \textbf{8}, 639 (1968)

Antonova, A.E., Shabansky, V.P., Hedgecock, P.C.: Geomagn. Aeron. \textbf{23},
574 (1983)

Carpenter, L.R., Anderson, R.R.:\ J. Geophys. Res. \textbf{97}, 1097 (1992)

Chappel, C.R.:\ J. Geophys. Res. \textbf{79}, 1861 (1974)

Denton, R.E., Takahashi, K., Anderson, R.R., Wuest, M.P.: J. Geophys. Res.
\textbf{109}, A06202 (2004)

Denton, R.E., Takahashi, K., Galkin, I.A., Nsumei, P.A., Huang, X., Reinisch,
B.W., Anderson, R.R., Sleeper, M.K., Hughes, W.J.: J. Geophys. Res.
\textbf{111}, A04213 (2006)

Feygin, F.Z., Pokhotelov, O.A., Pokhotelov, D.O., Mursula, K., Kangas, J.,
Braysy, T., Kerttula, R.: J. Geophys. Res. \textbf{103}, 20.481 (1998)

Fraser, B.J., McPherron, R.L.: J. Geophys. Res. \textbf{87, }4560 (1982)

Fraser, B.J., Samson, J.C., Hu, Y.D., McPherron, R.L., Russel, C.T.: J.
Geophys. Res. \textbf{97, }3063 (1992)

Guglielmi, A.V., Pokhotelov, O.A., Stenflo, L., Shukla, P.K.: Astrophys. Space
Sci. \textbf{200, }91 (1993)

Guglielmi, A.V., Pokhotelov, O.A.: Space Sci. Rev. \textbf{65}, 5 (1994)

Guglielmi, A.V., Pokhotelov, O.A., Feygin, F.Z., Kurchashov, Yu.P., McKenzie,
J.F.,\ Shukla, P.K.,\ Stenflo, L., Potapov, A.S.: J. Geophys. Res.
\textbf{100}, 7997 (1995)

Kozyra, J.U. Cravens,T.E., Nagy, A.F., Fonthim, E.G.: J. Geophys. Res.
\textbf{89, }2217 (1984)

Landau and Lifshits: Electrodinamica sploshnykh sred. Nauka, Moscow. P. 185 (1982)

Lemaire, J.: Phys. Fluids B \textbf{1}, 1519 (1989)

Mauk, B.N., McIlwain C.E., McPherron, R.L.: Geophys. Res.Lett.,\textbf{\ 8, }103\ (1981)

Nekrasov, A.K., Feygin, F.Z.: Physica Scripta \textbf{71}, 310 (2005)

Nekrasov, A.K., Feygin, F.Z.: Ann. Geophys. \textbf{24}, 467 (2006)

Nekrasov, A.K., Feygin, F.Z.: Nonlin. Proc. Geophys. \textbf{18, }235
\textbf{(}2011\textbf{)}

Nekrasov, A.K., Feygin, F.Z.: Astrophys. Space Sci. \textbf{341}, 225
\textbf{(}2012\textbf{) }

Persoon, A.M., Gurnett, D.A., Santolik, O., Kurth, W.S., Faden, J.B., Groene,
J.B., Lewis, G.R., Coates, A.J., Wilson, R.J., Tokar, R.L., Wahlund, J.-E.,
Moncuquet, M.: J. Geophys. Res. \textbf{114}, A04211 (2009)

Pokhotelov, O.A., Feygin, F.Z., Stenflo, L., Shukla, P.K.: J. Geophys. Res.
\textbf{101}, 10.827 (1996)

Takahashi, K., Denton, R.E., Anderson, R.R., Hughes, W.J.: J. Geophys. Res.
\textbf{111}, A01201 (2006)

Tsyganenko, N.A.: J. Geophys. Res. \textbf{100}, 5599 (1995)

Witt, E., Hudson, M.K., Li, X., Roth, I., Temerin, M.: J. Geophys. Res.:
\textbf{100},\textit{\ }12.151 (1995)

Young, D.T., Perraut, S.., Roux, A., Villedary, C., Gendrin, R., Korth, A.,
Kremser, G, Jones, D.: J. Geophys. Res. \textbf{86,} 6755 (1981)

\bigskip

\section{Figures}

\begin{figure}
\begin{center}
      \includegraphics[width=0.98\textwidth]{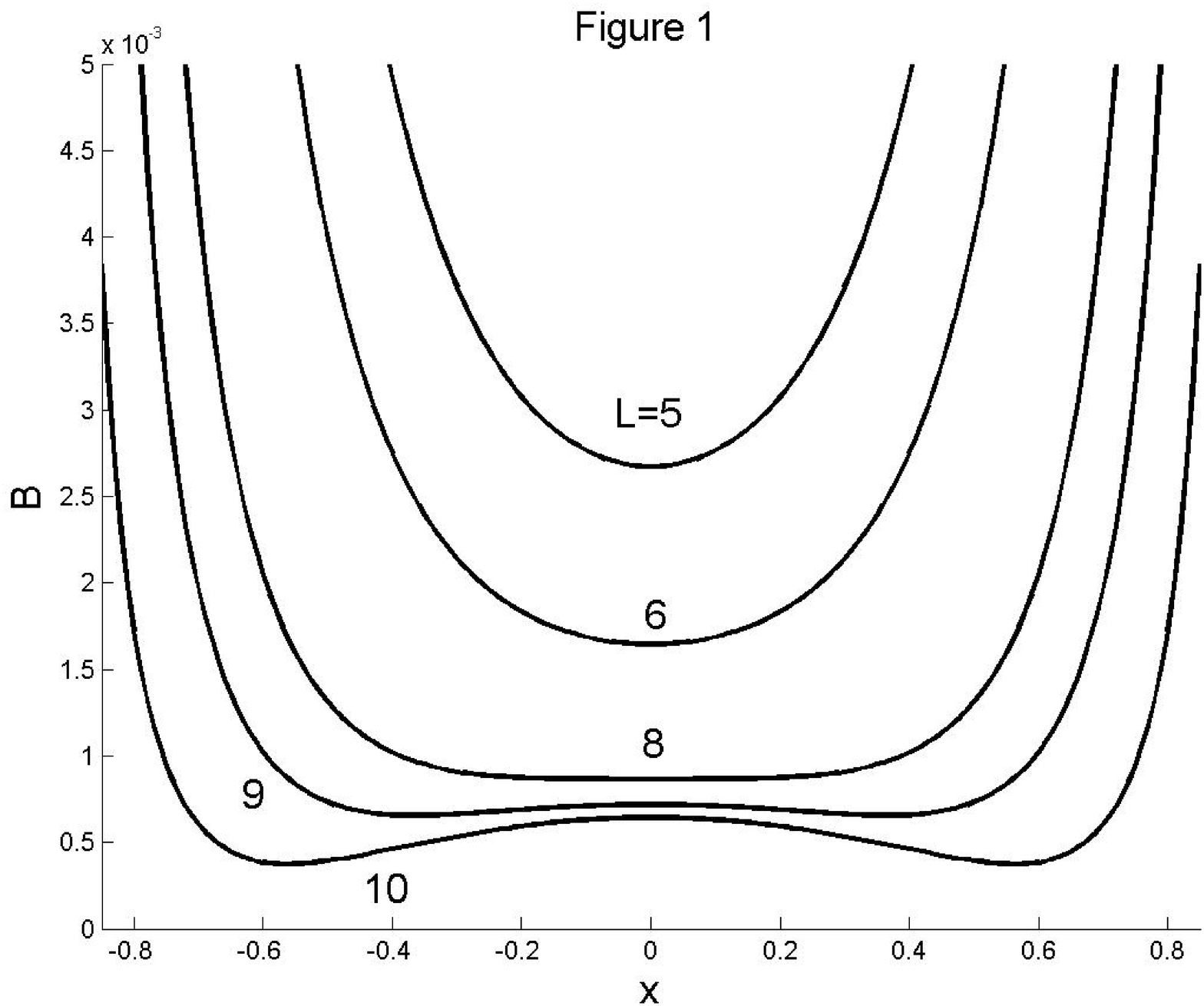}
 \caption{The dependence of the Earth's magnetic field $B$ (in G) on the
geomagnetic latitude $\varphi$ in the meridional plane of the dayside
magnetosphere for different $L$ in the model by Antonova and Shabansky (1968)}
\label{Fig:1}
 \end{center}
\end{figure}

\begin{figure}
\begin{center}
        \subfigure{\includegraphics[width=0.75\textwidth]{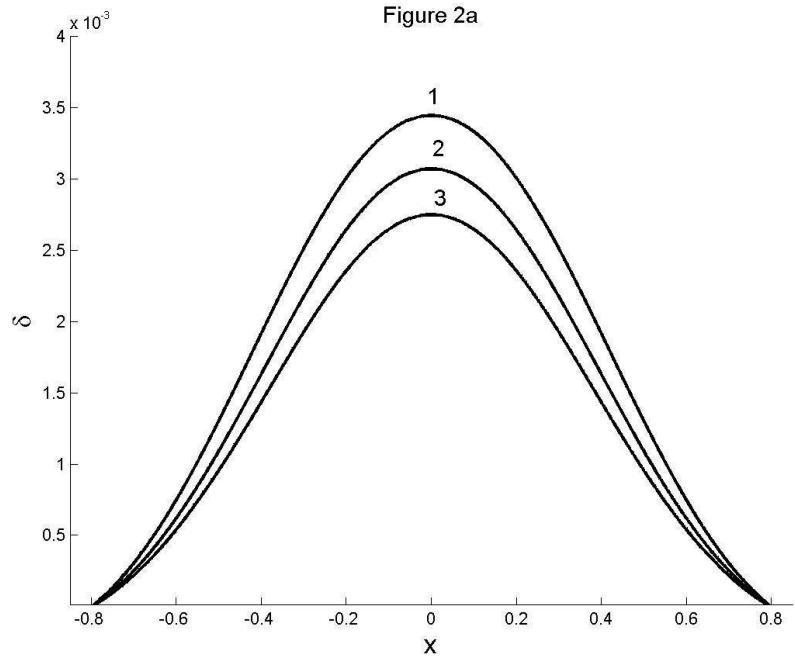}}
        \subfigure{\includegraphics[width=0.75\textwidth]{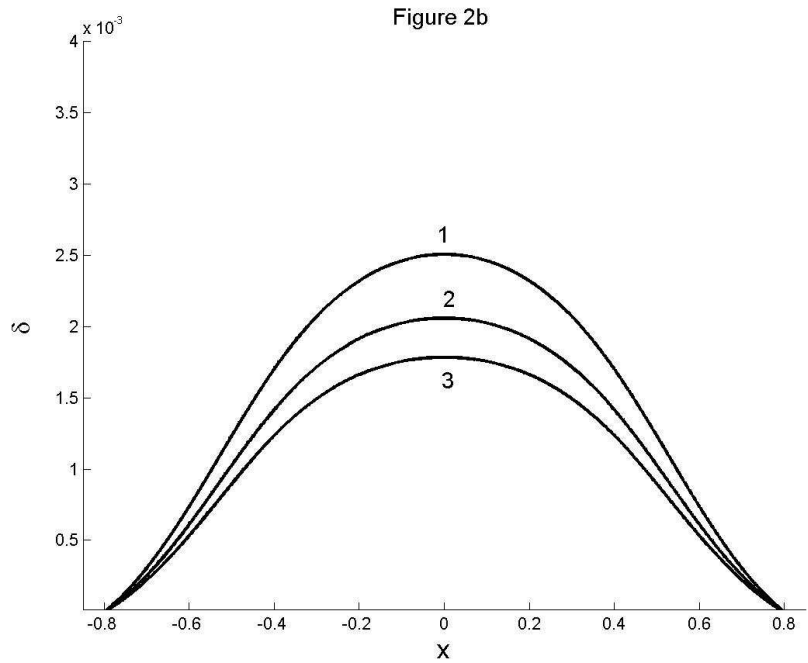}}
 \caption{a) The dependence of the relative density perturbation $\delta$ along the
field line at $L=6$ and $B_{10}=10^{-5}$ G. The curve $1$ corresponds to
$H_{0}=0$, curve $2$ - $H_{0}=0.5$, curve $3$ - $H_{0}=1.0$. b) The same as in Fig. 2a at $L=10$.}
\label{Fig:2}
 \end{center}
\end{figure}
%

\begin{figure}
\begin{center}
        \subfigure{\includegraphics[width=0.75\textwidth]{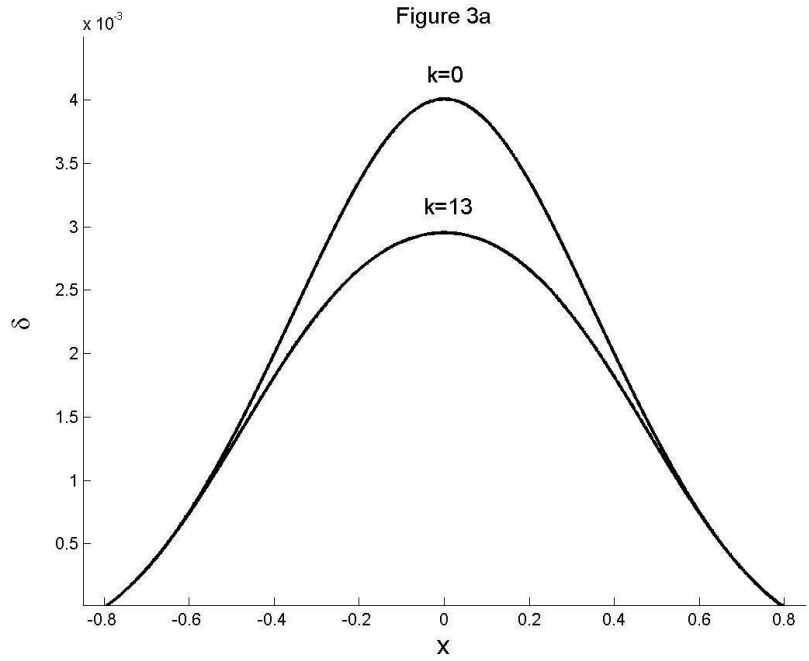}}
        \subfigure{\includegraphics[width=0.75\textwidth]{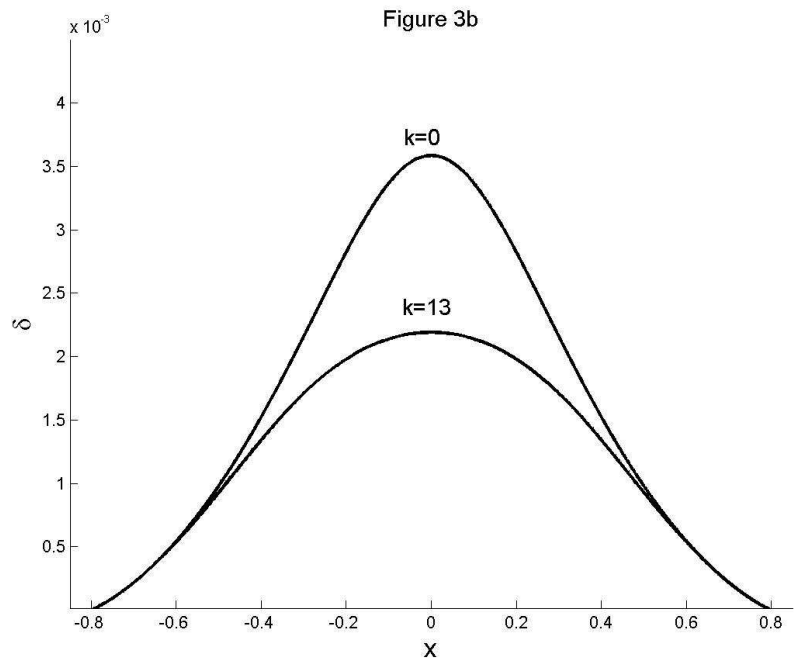}}
 \caption{a) The dependence of $\delta$-distribution on $x$ in the dipole ($k=0$)
and two-dipole ($k=13$) geomagnetic field at $L=8$ for $B_{10}=10^{-5}$ G and
$H_{0}=0$. b) The same as in Fig. 3a for $H_{0}=1.0$.}
\label{Fig:3}
 \end{center}
\end{figure}

%

\begin{figure}
\begin{center}
        \subfigure{\includegraphics[width=0.75\textwidth]{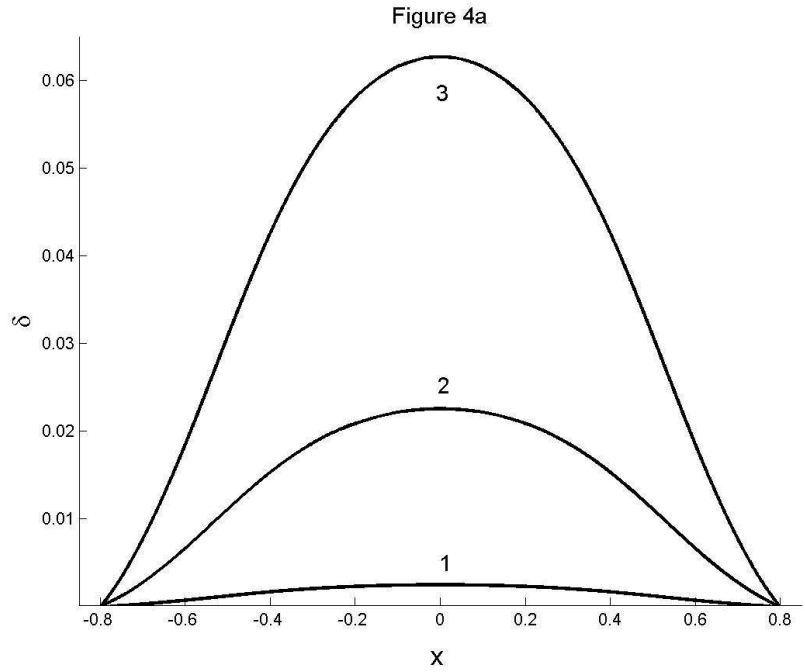}}
        \subfigure{\includegraphics[width=0.75\textwidth]{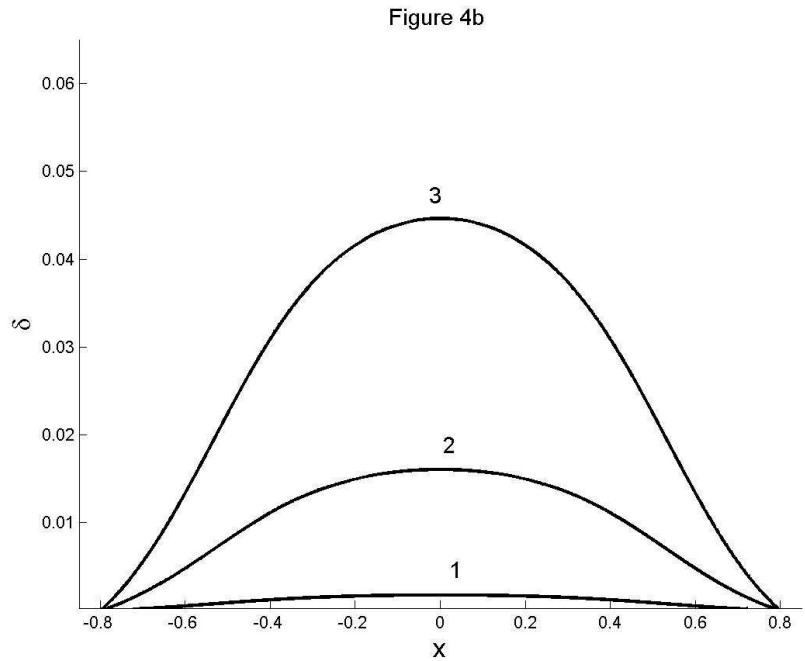}}
 \caption{a) The dependence of the relative density perturbation $\delta$ on $x$
for different values of the wave amplitude $B_{10}=(1,3,5)\times10^{-5}$ G at
$L=10$ and $H_{0}=0$. The curve $1$ corresponds to $B_{10}=10^{-5}$ G, curve
$2$ - $B_{10}=3\times10^{-5}$ G, curve $3$ - $B_{10}=5\times10^{-5}$ G. b) The same as in Fig. 4a for $H_{0}=1.0$.}
\label{Fig:4}
 \end{center}
\end{figure}

%

\begin{figure}
\begin{center}
      \includegraphics[width=0.98\textwidth]{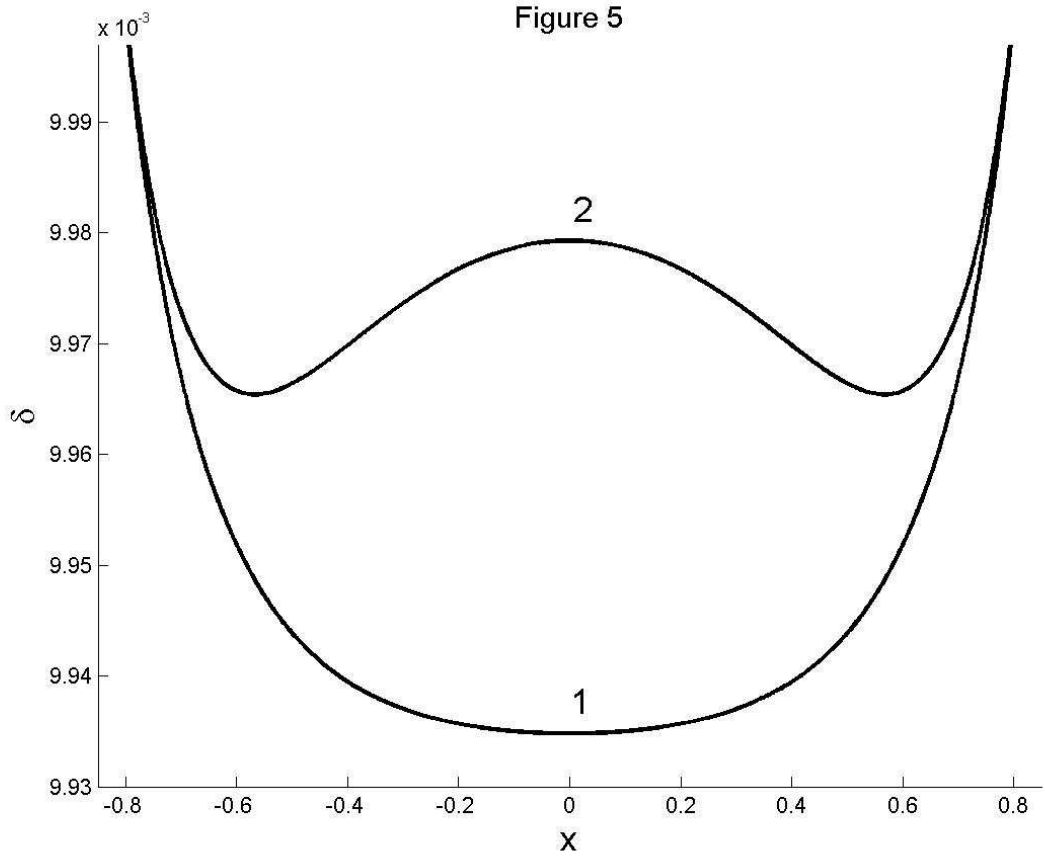}
 \caption{The dependence of $\delta$ on $x$ at $L=10$ and $B_{10}=0$ for
$H_{0}=0.5$. The curve 1 corresponds to $\Omega=0$. The curve 2 - $g_{eff}%
\neq0$.}
\label{Fig:5}
 \end{center}
\end{figure}

\bigskip
\end{document}